\documentclass{eas}
\usepackage{graphics}
\input{epsf}
\def\la{\hbox{\raise.35ex\rlap{$<$}\lower.6ex\hbox{$\sim$}\ }}
\def\ga{\hbox{\raise.35ex\rlap{$>$}\lower.6ex\hbox{$\sim$}\ }}

\def\beq{\begin{equation}}
\def\eeq{\end{equation}}
\def\beqa{\begin{eqnarray}}
\def\eeqa{\end{eqnarray}}
\def\bseq{\begin{subequations}}
\def\eseq{\end{subequations}}

\def\order#1{{\cal O}\left({#1}\right)}
\newcommand{\sfrac}[2]{\mbox{$\frac{#1}{#2}$}}
\begin{document}

\title{the magneto-rotational instability near threshold: spatio-temporal amplitude equation and saturation}
\runningtitle{Regev:  MRI - amplitude equation \& saturation}
\author{Oded Regev}
\address{Department of Physics, Technion - Israel Institute of Technology\\
\email{regev@physics.technion.ac.il}}
\secondaddress{Department of Astronomy, Columbia University}

\begin{abstract}
We show, by means of a perturbative weakly nonlinear analysis,
that the axisymmetric magneto-rotational instability (MRI) in a
magnetic Taylor-Couette (mTC) flow in a thin-gap gives rise,
for very small magnetic Prandtl numbers (${\cal
P}_{\rm m}$), to a real Ginzburg-Landau equation for the
disturbance amplitude. The saturation amplitude $A_s$ is found to scale
in this regime as  ${\cal P}_{\rm m}^\delta$, with $1/2<\delta<2/3$
(depending on the boundary conditions adopted).
The asymptotic results are shown to comply with
numerical calculations performed by using a spectral code. They suggest that
the transport due to the nonlinearly developed MRI may be
vanishingly small for ${\cal P}_{\rm m} \ll 1$.
\end{abstract}
\maketitle

\section{Introduction}
\subsection{General}
Asymptotic approaches to nonlinear stellar pulsation were pioneered by
Robert Buchler and his collaborators (see this volume).
Amplitude equations, which capture the essential dynamics
near the instability threshold, result from such approaches
and may usually be derived using singular perturbation theory (see Marie-Jo
Goupil, this volume). In this contribution we report on the results of applying a
perturbative asymptotic analysis to an instability that has acquired, in the
past 15 years or so, a paramount importance in the quest to understand angular
momentum transport in accretion disks, namely the magneto-rotational instability
(MRI). Even though the system studied here is rather different from
pulsating stars, the ideas and techniques are quite similar. One important difference
is that while the amplitude equations for stellar pulsation are sets
of ordinary differential equations, here we must take recourse to partial differential
equation(s) (PDE), in which the amplitude (usually called in this case {\em envelope}) is
a function of time (as in the ordinary amplitude equations) {\em and} space.
The application of singular perturbation theory requires thus the use of slow
variation in time as well as in space. In this way spatio-temporal slow dynamics
(i.e., patterns) is captured by a generic PDE - the real Ginzburg-Landau Equation, rGLE,
in our case.

The rudiments of singular perturbation theory are well explained in the book by
Bender \& Orszag (\cite{BO}), while a detailed account on problems
dealing with the dynamics on the center manifold
can be found in Manneville (\cite{manbook}).
It seems that these powerful analytical and semi-analytical techniques have not yet found
their way to enough astrophysical applications. In the early 1980s Robert Buchler
started an ambitious program of introducing such a new approach to stellar pulsation.
I was fortunate to be his postdoc then, but have moved on to other topics. Robert
and his postdocs (most of which contributed to this volume) have pursued this program and
during this conference one could learn just how much has been achieved in the understanding
of stellar pulsation, following asymptotic approaches, combined with numerical and
phenomenological ones. It is quite unfortunate that "driving
forces" of Robert's magnitude have unfortunately been quite rare in other branches
of astrophysics.
\subsection{The MRI - background}
 The linear MRI has been known for almost 50 years
(Velikhov \cite{veli}, Chandrasekhar \cite{chandra1}):
Rayleigh stable rotating Couette flows
of conducting fluids are
destabilized in the presence of a vertical magnetic field
 if $d\Omega^2/dr<0$
(angular velocity decreasing outward). However, the MRI only
acquired importance to astrophysics after the influential work of
Balbus \& Hawley (\cite{baha1}), who demonstrated its viability to
{\em cylindrical} accretion disks (for any sufficiently weak field).
The linear analysis was later supplemented by nonlinear numerical simulations
(albeit of a small, shearing-box (SB), segment of the disk). Enhanced
transport (conceivably turbulent) is necessary for accretion to
proceed in disks, found in a variety of astrophysical settings -
from protostars to active galactic nuclei. As is well known, Keplerian
rotation law is hydrodynamically linearly stable and thus
the MRI has been widely accepted as an
attractive solution for enhanced transport (see the
reviews by Balbus \& Hawley \cite{rev1} and Balbus \cite{rev2}), even though
some questions on the nature of MRI-driven turbulence still remain (see, e.g.,
Branderburg \cite{brand}), in particular in view of some recent numerical
studies (Pessah, Chan \& Psaltis \cite{pcp07}, Fromang \& Papaloizou \cite{pap1},
Fromang {\em et} al. \cite{pap2}).
Because of the MRI's
importance and some of its outstanding unresolved issues, several
groups have quite recently undertaken projects to investigate the
instability under laboratory conditions. Additionally, numerical simulations
specially designed for experimental setups have been conducted. It appears, however,
quite impossible to make definite deductions
from these experiments and simulations (see, e.g., Ji {\em et} al. \cite{jgk},
Noguchi {\em et} al. \cite{colgate}, Sisan {\em et} al. \cite{sisan},
Stefani {\em et} al. \cite{rudiger}, and references therein).

We shall report here on the result of a weakly nonlinear
analysis of the MRI near threshold, for a dissipative mTC flow.
We have done the analysis for the above  simplified (relatively to an accretion disk)
problem for two types of boundary conditions
(see Umurhan, Menou \& Regev \cite{umr07}, hereafter UMR and Umurhan, Regev \& Menou
\cite{urm07}, hereafter URM, for a detailed account).
This kind of approach is important because the
viability of the MRI as the driver of
turbulence and angular-momentum transport
relies on understanding its
nonlinear development and saturation.
By complementing numerical simulations with
analytical methods useful physical insight can be expected. To
facilitate an analytical approach we made a number of simplifying
assumptions so as to make the system amenable to well-known
asymptotic methods (see, e.g., Cross and Hohenberg \cite{cross}, Regev \cite{mybook}) for the
derivation of a nonlinear envelope equation, valid near the linear instability threshold.
\section{Linear theory}
\subsection{Equations}
The hydromagnetic equations in cylindrical coordinates (Chandrasekhar \cite{chandra2})
are applied to the neighborhood of a representative radial point
($r_0$) in a mTC setup with an imposed constant background vertical magnetic
field.  The steady base flow has only a
constant vertical magnetic field, ${\bf B} = B_0 {\bf \hat z}$, and a
velocity of the form ${\bf V}=U(x){\bf {\hat y}} $. In this base state
the velocity has a linear shear profile $U(x)=-q\Omega_0 x$,
representing an azimuthal flow about a point $r_0$, that rotates with
a rate $\Omega_0$, defined from the differential rotation law
$\Omega(r) \propto \Omega_0 (r/r_0)^{-q}$ ($q=3/2$ for Keplerian rotation).
The total pressure in the
base state (divided by the constant density),
$ \Pi \equiv {\rho_0}^{-1}\left(P+\frac{B_0^2}{8\pi}\right)$,
is a constant and thus its gradient is zero.

This base flow is disturbed by 3-D perturbations on the magnetic field
${\bf b}= (b_x,b_y,b_z)$, as well as on the velocity - ${\bf
u}=(u_x,u_y,u_z)$, and on the total pressure - $\varpi$.  We consider
only axisymmetric disturbances, i.e. perturbations with structure only
in the $x$ and $z$ directions.  This results, after
non-dimensionalization, in the following set of
equations, given here in the rotating frame:
\beqa
\frac{d{\bf u}}{dt} -2{ {\Omega_0 \bf{ \hat
z}}\times{\bf u}} -{q\Omega_0 u_x {\bf{\hat y}}} - {\cal C} {\bf
b}\cdot\nabla{\bf b} -{\cal C}B_0\partial_z {\bf b} &=& -\nabla\varpi
+\frac{1}{{\cal R}}\nabla^2 {\bf u}, \\ \frac{d{\bf b}}{dt} - {\bf
b}\cdot\nabla{\bf u} + {q\Omega_0 b_x {\bf{\hat y}}} -B_0\partial_z
{\bf u} &=& \frac{1}{{\cal R}_{{ m}}}\nabla^2 {\bf b},
\eeqa
\beq
\nabla\cdot{\bf u} \equiv \partial_x u_x + \partial_z
u_z = 0, \qquad \nabla\cdot{\bf b} \equiv \partial_x b_x + \partial_z
b_z = 0.
\label{incompressible_conditions}
\eeq
The Cartesian coordinates $x,y,z$ correspond to the radial
(shear-wise), azimuthal (stream-wise) and vertical directions,
respectively, and since axisymmetry is assumed $\nabla \equiv {\bf \hat
x}\partial_x+ {\bf \hat z}\partial_z$ and  $\nabla^2
\equiv \partial_x^2 + \partial_z^2$.  Lengths have been
non-dimensionalized by $L$
($\approx$ the gap size), time by the
local rotation rate $\tilde\Omega_0$ (tildes denote dimensional
quantities).  Because the dimensional rotation rate of the box (about
the central object) is ${\bf \tilde \Omega_0} = \tilde \Omega_0 {\bf
\hat z}$, the non-dimensional quantity $\Omega_0$ is simply
equal to $1$, but we keep it to flag the Coriolis terms.
Velocities have been scaled by $\tilde \Omega_0 L$ and the magnetic
field by its background value  $\tilde B_0$.
Thus $B_0 \equiv 1$ as well, but
again, we leave it in the equation set for later convenience.
The hydrodynamic pressure is scaled by $\tilde\rho_0 L^2
\tilde \Omega_0^2$ and the magnetic one by $\tilde B_0^2/(8 \pi)$.
The non-dimensional perturbation $\varpi$ of the total pressure
divided by the density (which is equal to 1 in non-dimensional units),
which survives the spatial derivatives, is thus given by $ \varpi =
p + {\cal C} \sfrac{1}{2} |{\bf b}|^2$, where $p$ is the
hydrodynamic pressure perturbation. The non-dimensional
parameter ${\cal C} \equiv \tilde B_0^2/(4\pi\tilde
\rho_0\tilde\Omega_0^2L^2) = \tilde V_A^2/\tilde V^2$
is the {\em Cowling number}, measuring the relative importance of the
magnetic pressure to the hydrodynamical one. It is equal to the
inverse square of the typical Alfv\'en number ($\tilde V_A$ is the
typical Alfv\'en speed).  The Cowling number appears in the non-linear
equations, together with the two {\em Reynolds numbers}:
${\cal R} \equiv {\tilde\Omega_0 L^2}/{\tilde \nu}$ and
${\cal R}_m \equiv {\tilde\Omega_0 L^2}/{\tilde \eta}$,
where $\tilde \nu$ and $\tilde \eta$ are, respectively, the
microscopic viscosity and magnetic resistivity of the fluid.  We shall
also see that the {\em magnetic Prandtl number}, given as ${\cal P}_m
\equiv {\cal R}_{ m}/{\cal R}$, plays an important role in the
nonlinear evolution of this system.

It is useful to rewrite the above equations of motion in terms of more convenient
dependent variables. Because the flow is incompressible and $y$-independent, the radial and
vertical velocities can be expressed in terms of the {\em
streamfunction}, $\Psi$, that is, $(u{_x},u{_z}) =
(\partial_z\Psi,-\partial_x \Psi)$. Also, since the magnetic field is
source free, one can similarly express its vertical and radial components
in terms of the {\em flux function}, $\Phi$, that is, $(b{_x},b{_z})
=(\partial_z\Phi,-\partial_x \Phi)$ (see URM for details).
The system is supplement by appropriate boundary condition on the thin
cylindrical gap "walls",
as detailed in UMR and URM and we also point out that the stress relevant for
angular momentum transport, $\Sigma$, is
\[
{\Sigma} ={\Sigma}_{\rm R}+{\Sigma}_{\rm M},\qquad
{\Sigma}_{\rm R} \equiv
u_x u_y, \quad {\Sigma}_{\rm M} \equiv -{\cal C} b_x b_y.
\]
which consists of
${\Sigma}_{\rm R}$ and ${\Sigma}_{\rm M}$, the Reynolds
(hydrodynamic) and Maxwell stresses, expressing the velocity and
magnetic field disturbance correlations, respectively.
\subsection{Linear stability}
Linearization of (2.1-2.3), for perturbations of the
form $\propto e^{st + ik_x x + ik_z z}$, gives rise to the dispersion relation,
\[
{\cal D}(s;k_x,k_z,{\cal P}_{\rm m},{\cal S},{\cal C},q)
=a_0s^4 + a_1s^3 + a_2s^2 + a_3s + a_4 = 0,
\]
where all the $a_i$ are functions of $k_x$, $k_z$ and the other parameters, but we shall write
explicitly only the coefficient that will be used in what follows,
\beq
a_4 =
 \frac{{\cal C}}{{\cal S}^4}\Bigl[
k_{_T}^2{\cal C}({\cal C}k_{_T}^4 {\cal P}_{\rm m} + k_z^2 {\cal S}^2)^2
+\kappa^2 {\cal S}^2{\cal C} k_{_T}^4 k_z^2 -2 q   {\cal S}^4k_z^4
 \Bigr],
 \label{a_coefficients}
\eeq
where the notation $\kappa^2 \equiv
2(2-q), \ k_{_T}^2 \equiv k_x^2 + k_z^2$
is introduced.

For given values of the parameters (call them
$p_j$) there will be four distinct modes. The linear theory in
various limits for this problem has been discussed in numerous
publications and we shall not elaborate
upon it any further. Rather, we focus on situations where the
most unstable mode (of the four) is marginal (critical) for some
$k_z$ at a given value of $k_x = K$. We fix $K$ in order to
focus on the marginal vertical dynamics within the thin gap
(actually a rotating channel, see below).
Marginality ($s=0$) implies the vanishing
of the real coefficient $a_4(k_z)$ (as expressed above)
and its derivative with respect to $k_z$ at some $k_z=Q$:
\beq
a_4(k_z=Q; K, p_j)=0, \qquad \frac {\partial a_4}{\partial k_z}(k_z = Q; K, p_j)=0.
\eeq
\begin{figure}
\begin{center}
\leavevmode \epsfysize=5.0
cm
\epsfbox{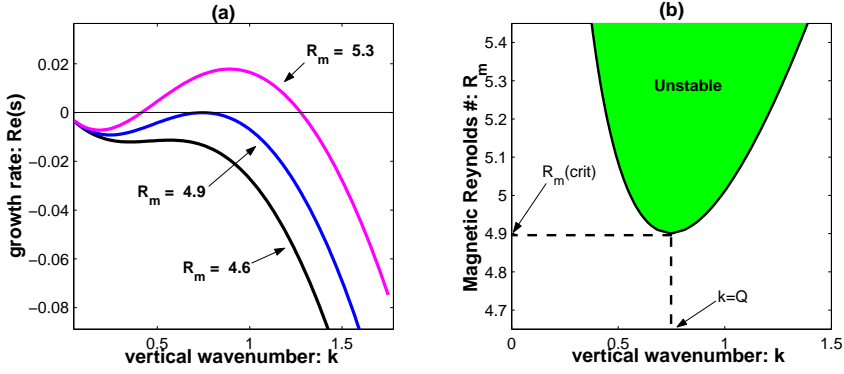}
\end{center}
\caption{{\small
Summary of linear theory.  This example is for ${\cal C} = 0.08$,
${\cal P}_m = 0.001$, $q = 3/2$, and the fundamental mode.
(a) Growth rates, $Re(s)$, as a function
of wavenumber $k$ for three values of ${\cal R}_{m}$.  (b) Solid line
depicts those values of ${\cal R}_{m}$ and $k$ where $Re(s)=0$.  The
shaded region shows unstable modes.  The locations of $k=k_{crit}\equiv Q$ and
${\cal R}_m= {\cal R}_{m}({\rm crit}) \equiv  R_m$ are shown.
}}
\label{linear_plottery_Pm001}
\end{figure}
The second condition and (\ref{a_coefficients}) yield
\beqa
&& Q\Bigl[Q^2{\cal S}^4({\cal C}Q^2 - 4q )
+(K^2 + Q^2)2Q^2{\cal C}{\cal S}^2({\cal S}^2+ \kappa^2)+ \nonumber \\
&&(K^2 + Q^2)^2{\cal C}{\cal S}^2(6{\cal C}{\cal P_{\rm
m}}Q^2 + \kappa^2)+\nonumber \\
&& 2(K^2 + Q^2)^3{\cal P_{\rm m}}{\cal C}^2{\cal S}^2
+5(K^2 + Q^2)^4{\cal P_{\rm m}}^2{\cal C}^3\Bigl] =0. \label{da4dkz}
\eeqa
This equation, together with $a_4(k_z=Q, K; p_i)=0$, can be solved
for ${\cal S}$
and $Q$.
The general expressions for ${\cal S}(K,{\cal P}_{\rm m},{\cal C},q)$
and $Q(K,{\cal P}_{\rm m},{\cal C},q)$ are lengthy
but their asymptotic forms, to $\order{{\cal P}_{\rm m}}$
(for ${\cal P}_{\rm m} \ll 1$), are simple,
\beq
{\cal S} = \frac{\sqrt{16 \ {\cal C}q(2-q)} K}{(2q-{\cal C}K^2)},\qquad
 Q^2
= K^2\frac{2q-{\cal C}K^2}{2q+{\cal C}K^2}
 \label{def_S}
\eeq
If ${\cal C} K^2 > 2q$ the solutions are not physically meaningful,
while
${\cal C} K^2 = 2q$ corresponds to the ideal MRI limit.
\par
Our system is a rotating thin channel, whose
walls are at $x=0,\pi/K$. All quantities are vertically periodic on
a scale $L_z$ commensurate with integer multiples of $2\pi/Q$. As
long as $L_z \gg 1/Q$, the limit of a vertically extended
system (and thus a continuous spectrum of vertical modes) may be
effected. Linear theory, as discussed above, is
summarized in Figure \ref{linear_plottery_Pm001}

\section{Weakly nonlinear analysis}
We follow the fluid into instability by tuning the vertical
background field away from the steady state, i.e.
$ B_0
\rightarrow 1 - \epsilon^2\tilde\lambda, $
where $\epsilon \ll 1$ and $\tilde \lambda$ is an $\order 1$ control parameter.
This means that we are now in a position to apply procedures of
singular perturbation theory to this problem,
by employing the multiple-scale (in $z$ and $t$) method (e.g,
Bender \& Orszag \cite{BO}). It facilitates,  by imposing suitable solvability
conditions at each expansion order of the calculation in order
to prevent a breakdown in the solutions, a derivation of an
envelope equation for the unstable mode.
Thus for any dependent fluid quantity $F(x,z,t)$ we assume
\beq
F(x,z,t) = \epsilon F_1(x,z,t) + \epsilon^2 F_2(x,z,t)  + \cdots.
\label{expansion_scheme}
\eeq
The fact that the $x$ and $z$ components
of the velocity and magnetic field perturbations can be derived from
a streamfunction, $\Psi(x,z)$ and magnetic flux function $\Phi(x,z)$ reduces
the number of relevant dependent variables $F$ to four ($u_y$ and $b_y$
are the additional two). These four dependent variables will also be used
in the spectral numerical calculation (see below).

For the lowest $\epsilon$ order of the equations, resulting from substituting
the expansions into the original PDE and collecting same order terms, we make
the Ansatz
$
F_1(x, z,t) = \hat F_1 \tilde A(\epsilon z,\epsilon^2 t)e^{iQ z}\sin K x
+ {\rm c.c.},
$
where $\hat F_1$ is a constant (according to the variable in question), and
where the envelope function $\tilde A$
(an arbitrary constant amplitude in linear theory)
is now allowed to have weak space (on scale $Z\equiv \epsilon z$)
and time (on scale $T \equiv \epsilon^2 t$) dependencies.
Because
this system is tenth order in $x$-derivatives,
a sufficient number of conditions must be specified at the edges.
and the Ansatz has to obey them. In UMR we chose mathematically
expedient boundary conditions, which allowed fully analytical
treatment for the limit ${\cal P}_{\rm m}\ll 1$ and here we shall
report, in some detail, only on the results of that work.
We have also tried some different boundary
conditions (see URM) and in that case the rGLE coefficients had to be calculated
numerically. The qualitative behavior of the saturation amplitude was found to
be quite similar (see below).

\subsection{Real Ginzubrg-Landau equation}
The end result of the asymptotic procedure procedure is the well-known
real Ginzburg-Landau Equation (rGLE) which, for
${\cal P}_{\rm m} \ll 1$, is
\beq
\partial_T A = \lambda A
- \frac{1}{{\cal P}_{\rm m}{\cal C}} |A|A^2 + D \partial_{_Z}^2 A .
\label{landau_eqn}
\eeq
We shall consider only the
magnitude (real part) of $A$. In a 1D gradient system like this one
phase dynamics can only give rise to wavelength modulations and consequently
our main results are in no way affected by it.
The real envelope is $A \equiv {\sqrt \xi} \tilde A $,
$\lambda \equiv \zeta \tilde\lambda $, and the coefficients, for $q=3/2$  are

\beqa \xi &=&
\frac{3}{4}\cdot\frac{
5{\cal S}^4 - 18{\cal S}^2 - 32
+ 2({\cal S}^2 + 16)\sqrt{{\cal S}^2 + 1}}
{{\cal S}({\cal S}^2 +1)(4\sqrt{{\cal S}^2 +1}-3)}
, \ \ \ \\
D &=& 6\frac{\left({\cal S}^2 +2 - 2\sqrt{{\cal S}^2+1}\right)({\cal
S}^2 + 1)} {{\cal S}^3(4\sqrt{{\cal S}^2+1} - 3)},\\
\zeta &=& \frac{3{\cal S}\sqrt{{\cal S}^2 +1} - 6{\cal S}}
{4{\cal S}^2 +1 +\sqrt{{\cal S}^2+1}},
\label{landau_coefficients}
\eeqa
where ${\cal S}(K)$ is as given in (\ref{def_S}).
For ${\cal S} \gg 1$, (i.e. as one approaches the ideal MRI
limit), these
simplify to $ \xi =15/16, \ \zeta =3/4, \
D = 3/2$; and in general they remain $\order 1$ quantities for all
reasonable values of ${\cal S}$.

\subsection{Scaling of angular momentum transport}
Contributions to the
total angular momentum transport ($\dot {\rm J}= \dot {\rm J}_{\rm H} +
\dot {\rm J}_{\rm B}$)
due to the hydrodynamic
($\dot {\rm J}_{\rm H}$) and magnetic
correlations ($\dot {\rm J}_{\rm B}$) are,
to leading order,

\beqa
\dot {\rm J}_{\rm H} &=& \frac{9\epsilon^2}{{\cal S}}\left(
\frac{2+{\cal S}^2 - 2\sqrt{1+{\cal S}^2}}{1+{\cal S}^2}
\right)A^2
+\order{\epsilon^3,\epsilon^2{\cal P}_{\rm m}}, \ \ \ \ \
\nonumber
\\
\dot {\rm J}_{\rm B} &=& \frac{3\epsilon^2}{{\cal S}}\left(
1
-\frac{1}{\sqrt{1+{\cal S}^2}}\right)A^2
+\order{\epsilon^3,\epsilon^2{\cal P}_{\rm m}}.
\label{angular_momentum_flux}
\eeqa

The envelope $A$ is found by solving the  rGLE, which is a
well-studied system (see, e.g., Regev \cite{mybook}, for a summary and references).
It has 3 steady uniform solutions in 1D (here, the vertical):
an unstable state $A = 0$, and 2 stable states $A=\pm A_s$,
where $ A_s^2 = |\zeta({\cal S})|{\cal P}_{\rm m}{\cal C}$ (note that
$\zeta({\cal S})$ is an $\order{1}$ quantity).
$A_s$ is also the saturation amplitude,
because the system will develop towards it.

Setting $A \rightarrow A_s$ in
(\ref{angular_momentum_flux}) and in the expression for ${\rm E_V}$ (the total disturbance
energy, see UMR), followed by some algebra,
reveals that the angular momentum flux in the saturated state is, to leading order in
${\cal P}_{\rm m}$ and $\epsilon$,
$\dot{\rm J} =  \epsilon^2 |\zeta({\cal S})|
\gamma({\cal S}){\cal P}_{\rm m} {\cal C}{\cal S}^{-1}$,
while in the expression for the energy one term is independent
of ${\cal P}_{\rm m}$,
${\rm E_V}=\epsilon^4 {\cal C}^3 \zeta^2({\cal S}) \beta_2({\cal S}) +\order{{\cal P}_{\rm m}}$.
The key results of this analysis (see URM) are the following scalings (with the magnetic
Prandtl number)
\beq
A_s \sim \sqrt{\cal P}_{\rm m} \longrightarrow
{\rm E_V}\sim E_0,~~~~  \dot{\rm J}\sim {\cal P}_{\rm m}
\eeq
for ${\cal P}_{\rm m}\ll 1$,
where $E_0$ is a constant, independent of ${\cal P}_{\rm m}$
\footnote{For the boundary conditions used in URM we numerically got $A_s \propto {\cal P}_{\rm m}^{2/3}$}.
For fixed resistivity this implies $\dot{\rm J}\sim {\cal R}^{-1}$.

\begin{figure}
\begin{center}
\leavevmode \epsfysize=5.5cm
\epsfbox{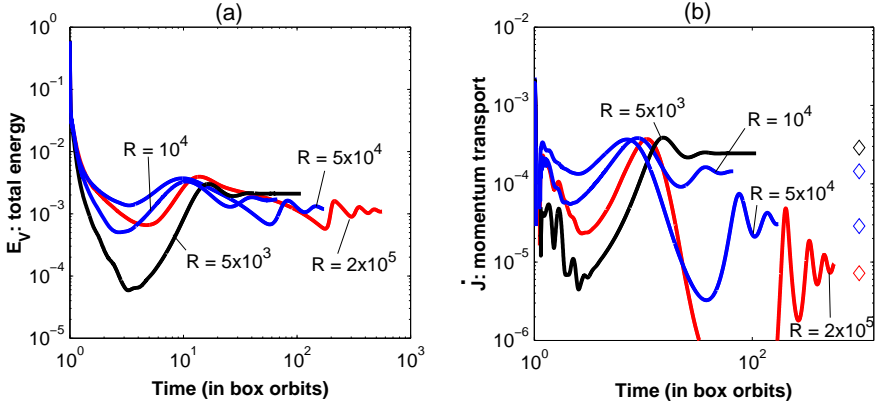}
\end{center}
\caption{{\small
${\rm E_V}$ (panel a) and $\dot {\rm J}$ (panel b) as a function of time
from numerical calculation. The different lines are labeled by the value
of the Reynolds number ${\cal R}$
The diamonds in panel (b) show the scaling predicted by our asymptotic
analysis, predicting also
a constant final value of the disturbance energy, as apparent in panel (a).
One "box orbit" $=2\pi$ in nondimensional units.
}}
\label{numerics_plots}
\end{figure}

We have also performed fully numerical calculations, using a
2-D spectral code to solve the original
nonlinear equations, in the streamfunction and magnetic flux
function formulation, near MRI threshold.
The code implemented a Fourier-Galerkin expansion
in ${\rm 64}\times{\rm 64}$ modes in each of
the four independent physical variables, i.e.
${\bf F}= \sum_{n,m}{\bf F}_{\rm nm}(t)
\sin K_n x e^{iQ_mz} + {\rm c.c.}$,
where ${\bf F} = (\psi,u_y,\Phi,b_y)^{{\rm T}}$
and where
${\bf F}_{\rm nm}(t)$ is the time-dependent amplitude
of the particular Fourier-Galerkin mode in question
(denoted by the indices $n,m$).
We typically started with
white noise initial conditions
on ${\bf F}_{\rm nm}(0)$
 at a level of 0.1
the energy of the background shear.
Because of space limitations we display
only a representative plot of runs made with
${\cal S} = 5.0$, ${\cal C} = 0.05$, (i.e. ${\cal R}_{\rm m}$ fixed),
$q=3/2$
and $\epsilon^2 = 0.2$ for a few successively increasing
values of ${\cal R}$ (Figure \ref{numerics_plots}).
With this value of $\epsilon$ the fastest growing linear mode has a
growth rate of $\sim 0.065$, and thus fully developed ideal MRI can not be
expected.
${\rm E}_{\rm V}$ saturates at a constant independent of
${\cal R}$ (for small enough ${\cal P}_{\rm m}$),
while $\dot{\rm J}$ saturates at values that scale as ${\cal R}^{-1}$. Thus
the asymptotic analysis fits very well the fully numerical results after the system
saturates.

\section{Summary}
The numerical and asymptotic solutions developed here
show that for ${\cal P}_{\rm m}
\ll 1$, it is the azimuthal velocity perturbation,
arising from the $\order {\epsilon^2}$ term, that becomes
dominant in the saturated state.
It appears to be the primary agent in the nonlinear saturation
of the MRI in the channel, acting anisotropically so as
to modify the shear profile and results in a
non-diagonal stress component (relevant for angular momentum transport).
Our analysis is complementary to the study of
Knobloch and Julien (\cite{kj}), who have performed
an asymptotic MRI analysis, but for a developed state, far from marginality.
\par
The trends predicted by this simplified model
are not qualitatively altered by different boundary conditions,
the value of $\delta$
changing only by a small amount.
Qualitatively similar scalings, but with
somewhat different values of $\delta>0$,
have very recently been reported by Lesur \& Longaretti (\cite{ll}) and Fromang {\em et} al.
(\cite{pap1}), however in those SB simulations ${\cal P}_{\rm m}$ was not taken to
be vanishingly small. It should be remarked that conventional accretion disks do have
${\cal P}_{\rm m}\ll 1$, but the numerical SB simulations cannot faithfully treat very small
such Prandtl numbers, for numerical reasons.
Further analytical investigations of the kind reported here should contribute
to better physical understanding of the MRI saturation..

\end{document}